# Hydration-controlled twist forms a moiré glass in charge-frustrated layered silicates

Juhyeok Lee[1,2], Piotr Zarzycki[3], Colin Ophus[4], Jim Ciston[2], Benjamin Gilbert[3,5], Jillian F. Banfield[3,5], Mary C. Scott[2,6], Michael L. Whittaker[1,3*]

1. Materials Sciences Division, Lawrence Berkeley National Laboratory, Berkeley, CA 94720
2. National Center for Electron Microscopy, Molecular Foundry, Lawrence Berkeley National Laboratory, Berkeley, CA 94720
3. Energy Geosciences Division, Lawrence Berkeley National Laboratory, Berkeley, CA 94720
4. Department of Materials Science and Engineering, Stanford University, Stanford, CA 94305
5. Department of Earth and Planetary Science, University of California, Berkeley, Berkeley, CA 94720
6. Department of Materials Science and Engineering, University of California, Berkeley, Berkeley, CA 94720

*Corresponding author. email: mwhittaker@lbl.gov

## Abstract

Twisting layered materials produces moiré superlattices, but prescribed twist angles are usually obtained by demanding assembly procedures. Here we show that montmorillonite, an abundant swelling clay, forms tunable moiré superlattices naturally. Focal-series high-resolution transmission electron microscopy, geometric phase analysis, and molecular dynamics simulation reveal that its apparent rotational disorder is biased toward low-angle misorientations inherited from discrete hydration states. Multilayer stacks preferentially adopt twists near 1–2°, 4°, and 10°, producing long-wavelength moirés without long-range rotational order. We define this kinetically trapped state as a *moiré glass*, distinct from featureless turbostratic stacking. Simulations indicate that lattice-charge disorder stabilizes the angular preferences, whereas charge ordering promotes random stacking. Hydration screens interlayer interactions and lubricates twist, while dehydration arrests the resulting configurations in discrete steps. These results establish dynamic hydration as a macroscopic handle for programming twist in layered matter.

# Main

Rotational misalignment between stacked layers is increasingly recognized as a key source of emergent behavior in layered materials[1–5]. In strongly interacting systems such as those with lattice charge, small twist angles generate moiré superlattices with large periodicities that spatially modulate strain, electrostatic potential, and interlayer chemistry, producing the coexistence of locally ordered and frustrated regions[4,6–10]. Although such moiré twist motifs are widespread in layered materials, including silicate clays[11,12], layered double hydroxides[13,14], MXenes[15,16], two-dimensional (2D) water ice[17], and graphene oxide[18–20], the physical origin and consequences of twist in these materials remain unresolved[21,22]. In particular, it is unclear whether rotational disorder reflects random stacking or instead corresponds to energetically structured preferred orientations[23–25].

A defining feature of charged layered materials is that layer rotation is often energetically costly, comparable to thermal, hydration, or shear energies[26–28]. As a result, rotational disorder is registry-dependent and tightly coupled to lattice strain, hydration, and chemical heterogeneity, giving rise to rugged energy landscapes with multiple metastable configurations[17,29]. This behavior contrasts with that of weakly interacting van der Waals bilayers, where barriers to twist are low and electronic band reconstruction gives rise to electronic moiré physics[1,28].

Montmorillonite (Mt) provides a particularly well-defined platform for interrogating rotational disorder among charged layers. Mt is commonly described as turbostratic: adjacent layers remain approximately parallel but lack long-range crystallographic registry, appearing disordered in transmission electron microscopy (TEM)[30,31] and bulk X-ray diffraction (XRD)[30,32]. This kind of stacking disorder is also widely reported in other strongly interacting layered materials[22,33–35]. Although turbostratic disorder is classically treated as random rotational stacking[21,36,37], experimental studies report small interlayer misorientations, typically below 10 degrees[30,38], suggesting that rotation is constrained by underlying energetics rather than being fully random. Such preferred misorientations are thought to arise from competition between heterogeneous intralayer electrostatics, set by the negative layer charge, and the organization of interlayer cations and hydration structures[12,39]. The resulting moiré structures in confined clays can in turn modulate interlayer ion distributions and exchange[39,40]. Despite these insights, a quantitative framework linking interlayer energetics, local stacking, structural heterogeneity, and preferred rotation angles is still lacking, because bulk-averaged probes obscure local, nearest-neighbor configurations.

Here, we investigate how rotational disorder emerges and persists in swelling clays by directly observing the rotational configuration space of thin Mt particles composed of two to ten layers. Using high-resolution transmission electron microscopy (HRTEM), combined with focal-series reconstruction, Fourier analysis, and geometric phase analysis (GPA)[41], we quantify both layer spacing and interlayer misorientation at the level of individual stacks and map the associated moiré domain deformation, strain localization, and bending- and tilt-induced distortions. These measurements are complemented by large-scale molecular dynamics (MD) simulations across variable hydration, and across random and ordered distributions of layer charge, which link the observed twist configurations to their underlying free-energy landscape. Together, these multiscale observations reveal that hydration enables rotational plasticity, whereas dehydration kinetically traps it, producing history-dependent stacking sequences. These sequences lack long-range rotational periodicity yet display preferred nearest-neighbor twist angles, which we trace to charge disorder. We refer to this nonequilibrium, structurally frustrated state as *moiré glass*, reflecting a kinetically trapped configuration within a rugged free-energy landscape. This framework provides a structural and energetic basis for understanding the multiscale mechanical and chemical response to hydration in charged layered materials.

## Results

### Statistical structure analysis using HRTEM images of Mt

Mt is a representative charged layered material comprising approximately 10 Å thick 2:1 layers (an octahedral sheet between two tetrahedral sheets) stacked along the *c*-axis and separated by hydrated interlayers (Fig. 1a). Isomorphic substitution within these sheets gives each layer a net negative charge that is compensated by exchangeable interlayer cations (Methods). The idealized model in Fig. 1a omits this substitution for clarity, but the MD simulations in Fig. 2 contain it.

Viewed along the *c*-axis ([001]), the single layer unit cell shows pronounced in-plane anisotropy, with *b* elongated relative to *a* (Fig. 1a). The three lattice planes with the largest interplanar spacings are (020), (110), and (–110) (Fig. 1b), and the angles between them deviate slightly from 60°, being compressed along *a* and elongated along *b* (Fig. 1c)[31]. Depending on the relative rotation of adjacent layers, these planes register to different degrees, producing projected stacking that ranges from crystalline through nearly crystalline to turbostratic.

To understand the rotational disorder in Mt, we first examined the structural features that can be directly measured from HRTEM images. Two parameters are particularly important for describing turbostratic superlattices: the number of layers within individual particles and the relative misorientation angle between adjacent layers (see Methods). The number of layers constrains the extent to which interlayer forces compete with intralayer elasticity, whereas the misorientation angle controls the formation of moiré periodicities that modulate cation ordering. Counting layers in edge view images (Fig. S1) shows that most particles exfoliated in clay-saturated aqueous phase contain only two to three layers (Fig. 1d). Fourier transform (FT) analysis of lattice images (Fig. 1f–m) yields a misorientation distribution that peaks near 1° and falls off rapidly up to 10° (Fig. 1e). Small rotations (e.g., 5°) produce long wavelength moiré lattices, whereas larger angles (>10°) give increasingly disordered interference. Because these moiré patterns create periodic domains of ordered and disordered interlayer species (cations and water), the strong preference for small angles indicates that misorientation is not random.

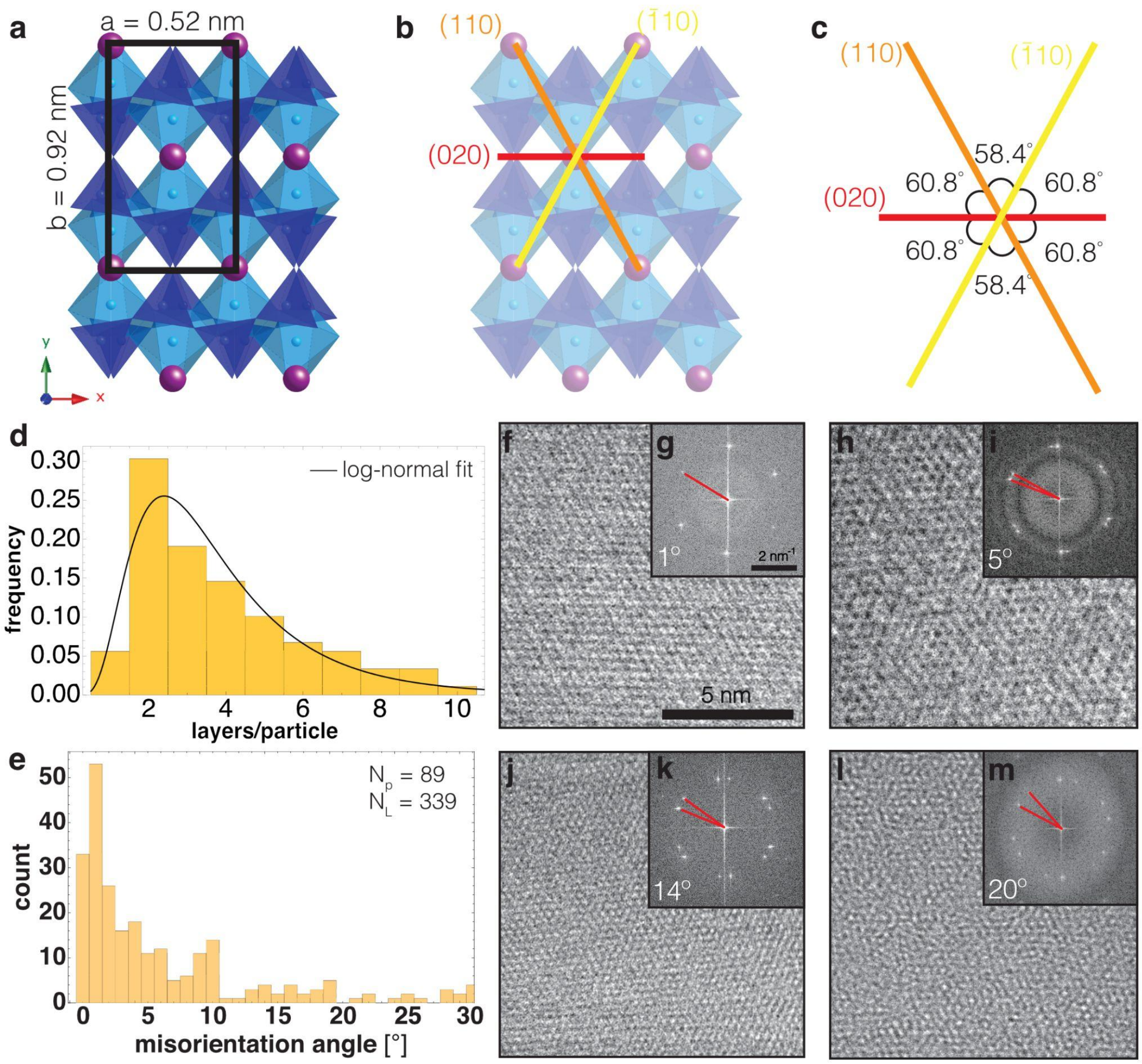

**Figure 1.** Mt particle size and orientation. (**a**) Unit cell of a single Mt layer viewed along the *c*-axis [001]. (**b**) Three lattice planes of Mt with the largest interplanar spacing (0.46 nm for (020) and 0.45 nm for (110) and (-110)). (**c**) Angular separation between lattice planes. Because $b > \sqrt{3}a$, the angles are slightly compressed in the *a* direction and elongated in the *b* direction. (**d**) Frequency distribution of the number of layers per particle. The solid line indicates fits to a lognormal distribution with a mean of 1.18 and a standard deviation of 0.56. (**e**) Histogram of relative misorientation angle between layers ($N_L$ is the number of layers, $N_P$ is the number of particles). (**f, h, j, l**) Low-dose high-resolution TEM images of four representative bilayer particles. (**g, i, k, m**) Respective FTs, showing layers separated by 1°, 5°, 14°, or 20°.

**Molecular dynamics simulation of the Mt bilayer**

We performed MD simulations of the Mt bilayer to calculate the misorientation-angle-dependent Gibbs free energies for systems containing one (1W), two (2W), and three (3W) intercalated water layers (see Fig. 2a-c and Methods). These free-energy profiles were then used to obtain the misorientation-angle probability distribution through Boltzmann weighting of the 1W, 2W, and 3W cases (see Fig. 2d-e and Methods), providing a basis for interpreting the observed misorientation preferences.

All three hydration states exhibit local free-energy minima at small misorientation angles near 1° and 2°, as well as at larger angles around 4°, 10°, and 14°. These minimum positions agree overall with the experimentally measured misorientation-angle histogram, whose dominant peak near 1° and weaker maxima at 4°, 10°, and 14° each fall on a calculated minimum (Figs. 1e, 2e). The observed misorientations are therefore not randomly distributed but sit at energy minima, with the population strongly concentrated at small angles.

Layers are hydrated when prepared from aqueous suspension and dried for TEM observation. The TEM operates under high vacuum, but we observe that the 1W state persists after one hour, and that the vacuum pressure is equivalent to that in the absence of a sample grid, indicating that the 1W state is stable under vacuum. Edge-view HRTEM images acquired under identical conditions systematically show a 1W interlayer spacing (Fig. S1), in agreement with XRD (Fig. S2) and with previous simulations[26] that find a 1W configuration under reduced hydration. However, the 1W state cannot by itself account for the observed small-angle preference; consistent with this, the Boltzmann weighting that reproduces the experiment is dominated by the more hydrated 2W and 3W energy landscapes (48% and 52%), with negligible 1W contribution. The distribution must therefore reflect the hydration history of the bilayer rather than its final dehydrated state.

The following sequence explains the observed populations. In the fully hydrated 3W state, the large interlayer spacing and lubricating water keep the layers mobile, and the global free-energy minimum lies at alignment (0°), so the bilayer equilibrates with most configurations near 0°. As it dehydrates, the 0° minimum is lost (it is absent in 2W) and rotational mobility falls, so these near-aligned configurations become kinetically trapped at small angles rather than relaxing further, which is the origin of the experimentally observed small-angle preference. Retaining just enough mobility to rotate slightly, the layers settle into the neighboring 1° minimum, which persists in the 2W and 1W states, and on full dehydration to 1W the bulk of the population is locked at this ~1° angle. The dominant ~1° peak thus reflects an energetic drive toward alignment in the hydrated bilayer, kinetically frozen into a small but finite twist during dehydration and producing the history-dependent distribution observed experimentally (Fig. 2e).

To test whether these small-angle wells originate from the random distribution of layer charge, we repeated the 1W calculation with the octahedral $Mg^{2+}$ substitutions placed on an ordered sublattice at equivalent total charge density (see Fig. S3 and Methods). The ordered configuration yields a smooth energy landscape that decreases nearly monotonically with angle, without the ~1–2° wells found in the disordered case. The locally preferred small-angle registries, and hence the rugged landscape underlying the moiré glass, therefore arise specifically from the random arrangement of the isomorphic substitution at specific lattice sites, rather than from the twist geometry at the layer scale. Physically, a slight rotational offset partially relieves the electrostatic frustration of this heterogeneous, quenched charge while preserving local registry, which stabilizes the small-angle wells.

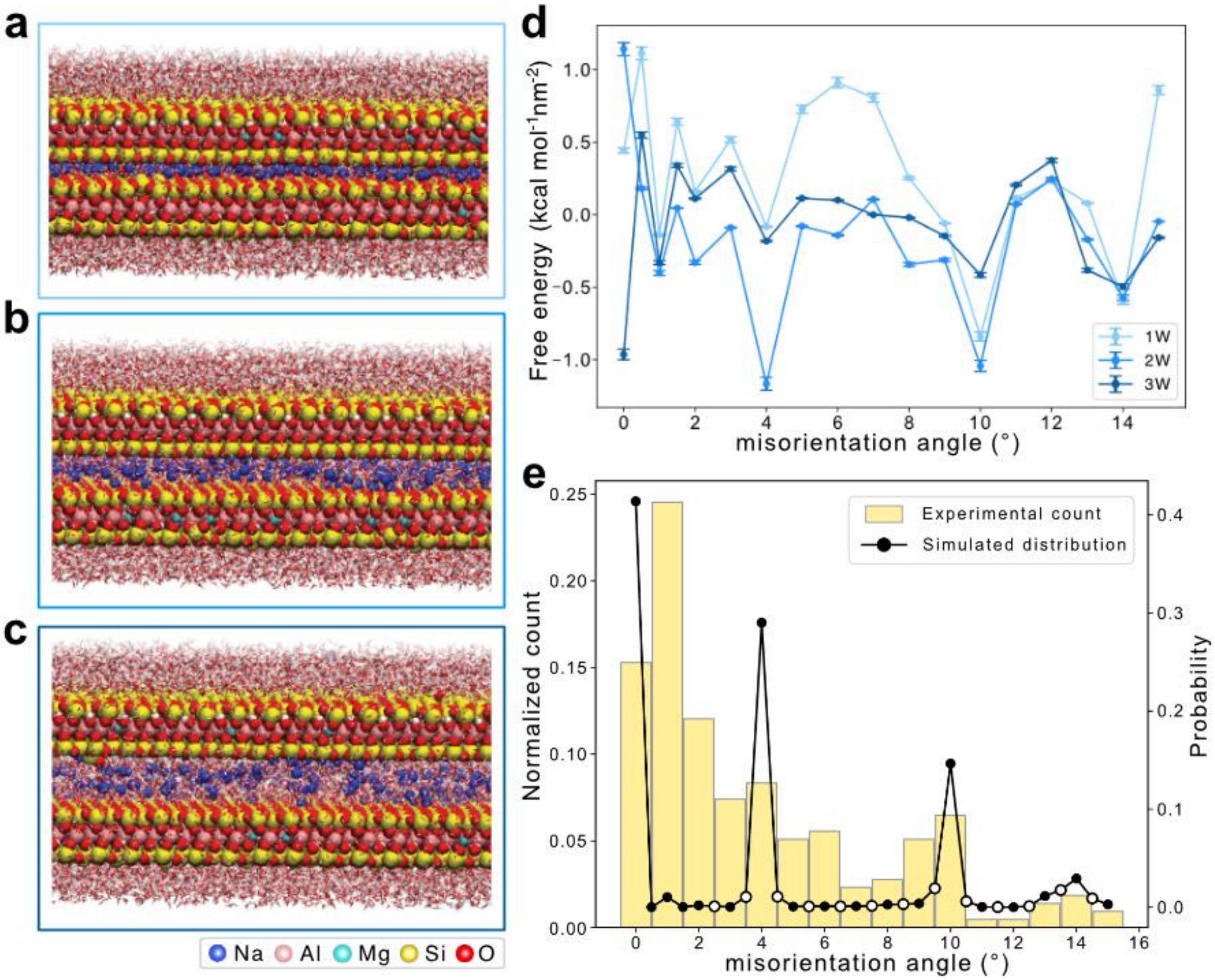

**Figure 2.** Comparison of observed and simulated misorientation-angle distribution in Mt bilayer. (**a–c**) Atomic structures of a bilayer Mt system with different numbers of intercalated water layers: one water (1W) layer (**a**), two water (2W) layers (**b**), and three water (3W) layers (**c**). Interlayer and surrounding water molecules are included and visualized as a diffuse distribution to represent the hydrated environment. (**d**) Calculated relative Gibbs free energy landscapes as a function of misorientation angle for bilayer Mt with 1W, 2W, and 3W layers. Free energies are shown relative to the minimum value for each hydration state. Error bars represent the root-mean-square fluctuations obtained from MD sampling and are small relative to the overall energy variation. (**e**) Experimental misorientation-angle distribution (yellow bars, normalized as in Fig. 1e) compared with the Boltzmann-weighted angular preference obtained from simulations by combining the 1W, 2W, and 3W cases (black symbols and line). The combined Boltzmann distribution consists of contributions of 0% from 1W, 48% from 2W, and 52% from 3W. The simulated distribution captures the preferential population at small misorientation angles observed experimentally. Open circles represent interpolated values obtained from the MD-simulated free-energy profiles prior to Boltzmann weighting.

**Focal series reconstruction of the Mt bilayer**

To resolve the local atomic structure underlying the dominant small-angle disorder, we performed focal series reconstruction on a representative bilayer with a misorientation of ~1° (see Methods). By reconstructing the exit wave phase, which is proportional to the projected atomic potential, this approach reaches an effective resolution of ~1 Å, substantially better than conventional HRTEM (Fig. S4). An overview of the reconstructed bilayer (Fig. 3a) shows extended regions with distinct lattice patterns, consistent with a slight rotational offset between the two layers.

Magnified views (Fig. 3b–e) reveal pronounced spatial heterogeneity in local interlayer registry. They include domains where the two layers align along all three principal directions (Fig. 3b), regions with partial registry along <110> (Fig. 3c) or along <020> and <1–10> (Fig. 3d), and areas with no apparent in plane alignment (Fig. 3e). The coexistence of these distinct registries within a single ~1° bilayer directly reflects moiré modulation of interlayer stacking and underscores the value of local, high resolution analysis beyond ensemble averaged measurements.

Fourier analysis of the reconstructed bilayer further illustrates these variations. The FT (Fig. 3f) shows a central hexagram arising from interlayer lattice strain, and magnified Bragg spots resolve a splitting of the (260) peaks by 0.84°, directly confirming the small interlayer misorientation. Together with the real-space images, these features show that even a nominally simple bilayer hosts a mosaic of locally ordered and disordered stacking motifs correlated with lattice orientation.

The ~1 Å resolution does not resolve individual atoms: the cluster-like features in Fig. 3b arise from the projected overlap of several atomic columns rather than isolated sites. To interpret the contrast, we built a bilayer model with ~0.8° misorientation, chosen to match the measured Bragg peak splitting, and computed its projected potential (see Methods). The simulated maps (Fig. S5) reproduce both extended moiré domains and the local ordered and disordered variations, including the hexagonal cluster features; the partially aligned configuration (Fig. S5g–i) reproduces the striped pattern of Fig. 3c,d, with apparent atomic positions traceable to the ideal model. The reconstructed phase also shows larger modulation amplitudes in aligned than in misaligned domains, consistent with enhanced potential contrast at favorable registries. The close agreement between experiment and simulation indicates that the experimental layer is well described by the ideal structure, with residual discrepancies attributable to local distortion or out of plane bending.

Bragg filtered images of the (110), (020), and (1–10) reflections (Fig. 3g–i) highlight the moiré domains as alternating ordered and diffuse regions. The (110) and (020) reflections (Fig. 3g,h) show clear stripe contrast indicative of aligned arrangements, whereas the (1–10) reflection (Fig. 3i) is largely disordered. Simulations from the ideal structure reproduce this anisotropy only when a global tilt is included: without tilt (Fig. S6), the hexagonal clusters and stripe patterns are absent, whereas a 7° tilt about [130] (Fig. S5) recovers them, including the anisotropic (1–10) contrast (see Methods). The simulations do not fully capture the irregular stripe patterns and localized features, indicating that additional local distortions, such as bending or mesoscale variations, also contribute. The Bragg filtered contrast thus combines intrinsic moiré modulation with structural distortion, consistent with the spatially heterogeneous rotational disorder in Mt bilayers.

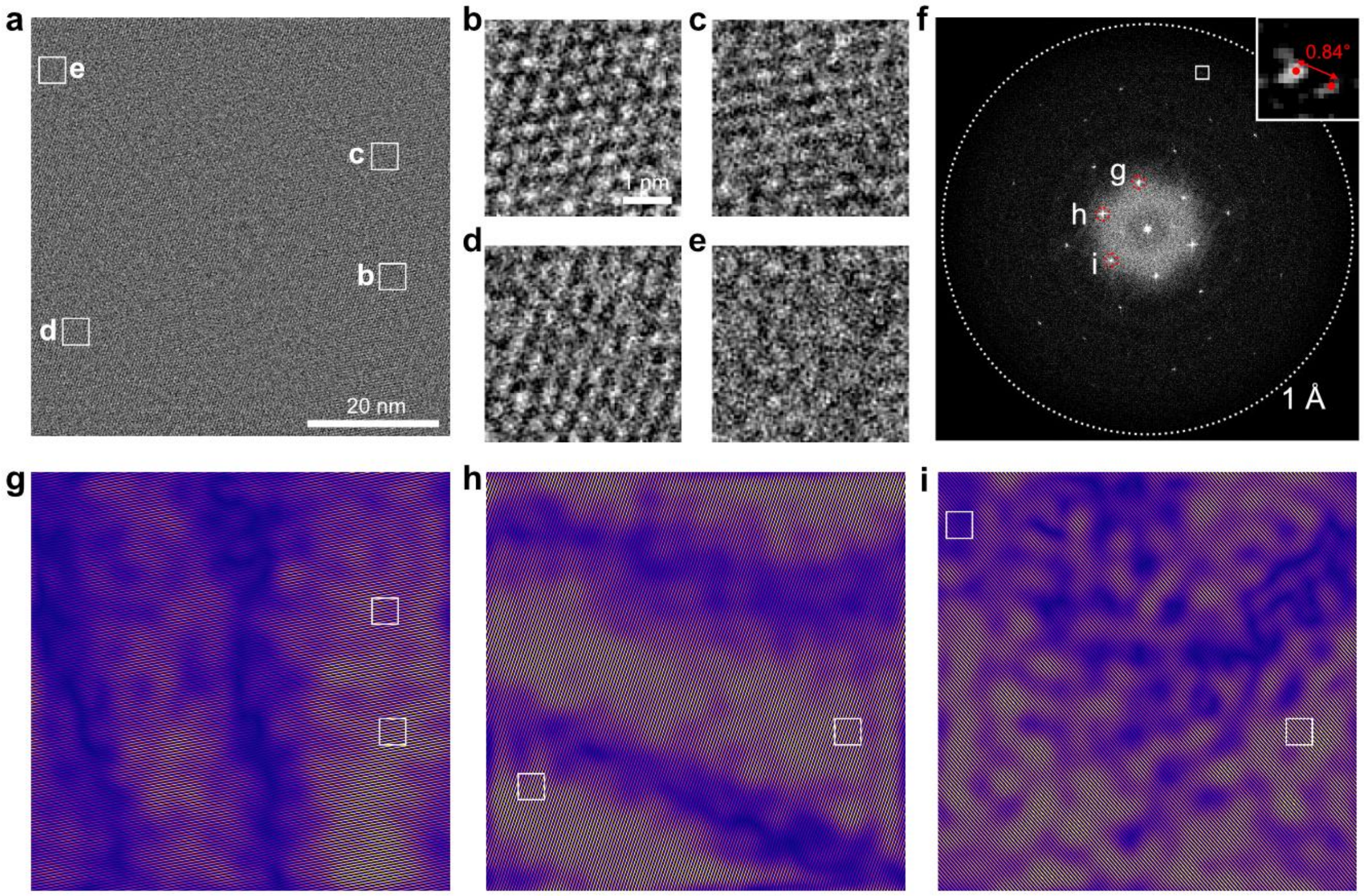

**Figure 3.** Focal series reconstruction and local stacking heterogeneity in an Mt bilayer. (**a**) Reconstructed phase image of a representative Mt bilayer exhibiting ~1° interlayer misorientation (scale bar, 20 nm). (**b-e**) Enlarged regions highlighting spatially heterogeneous interlayer registry, showing (**b**) alignment along all three principal directions, (**c**) alignment along <110> only, (**d**) alignment along (020) and (1-10), and (**e**) no alignment. (**f**) Fourier transform of (**a**), showing central hexagram pattern resulting from interlayer lattice strain. Enlarged Bragg spots from (**f**), revealing (260) peaks split by 0.84°. (**g-i**) Bragg filtered images from (**g**) (110) (**h**) (020), and (**i**) (1-10) overlaid on top of (**a**), revealing anisotropic contrast associated with local registry variations.

## Moiré distortion and layer-resolved strain analysis

We computed moiré distortion maps and rotation angles from the focal series reconstruction (Fig. 3) and the simulated potential (Fig. S5) by GPA (see Methods). Because the bilayer contains two misaligned lattices, GPA here uses references from the overlapping Bragg peaks of both layers, so the maps represent effective moiré distortion and rotation fields rather than true atomic-scale elastic strain within a single layer. The results comprise three distortion components (Exx, Eyy, Exy) and the in-plane rotation angle (Fig. 4).

The experimental maps show two prominent stripe features and several crossing points associated with the global interlayer rotation. These stripes arise from the long-wavelength moiré modulation produced by the relative twist and couple most strongly to the shear component Exy (Fig. 4c), whereas alternating contrast in Exx and Eyy (Fig. 4a,b) reflects periodic expansion and compression of the moiré lattice, that is, spatially varying accommodation of interlayer misfit set by the in-plane anisotropy of Mt. The simulated maps (Fig. 4e–h) reproduce the overall stripe features, although local magnitudes and signs differ, reflecting lattice relaxation, bending, and defects present in experiment but absent from the idealized model.

The experimental maps also contain localized dipole features. Although such dipoles could signal local out of plane distortion, similar dipoles along the (1–10) direction appear in the simulated maps, which include only interlayer rotation and a controlled global tilt; they therefore arise mainly from tilt induced phase variation rather than intrinsic lattice distortion. Overall, the effective fields show that rotational disorder generates nanoscale heterogeneity in the moiré structure, balancing interlayer strain accommodation against lattice coherence.

To separate these effective moiré fields from deformation within individual layers, we performed layer-resolved GPA using the well separated high order <330> reflections, which isolate Layer 1 and Layer 2 (see Methods; Fig. S7). Each layer shows pronounced anisotropic strain and rotation, but the two patterns differ in orientation and spatial distribution and show little spatial correlation, indicating that deformation is accommodated largely independently by each layer rather than shared. This weak coupling is consistent with the large interlayer separation of the hydrated (3W) state, with the independently accommodated patterns subsequently quenched into the collapsed (1W) configuration during nonequilibrium dehydration. These observations indicate that the long-wavelength moiré distortion fields observed in bilayer GPA arise from the superposition of distinct, layer-specific deformation patterns, linking moiré-scale modulation to heterogeneous lattice accommodation at the level of individual layers.

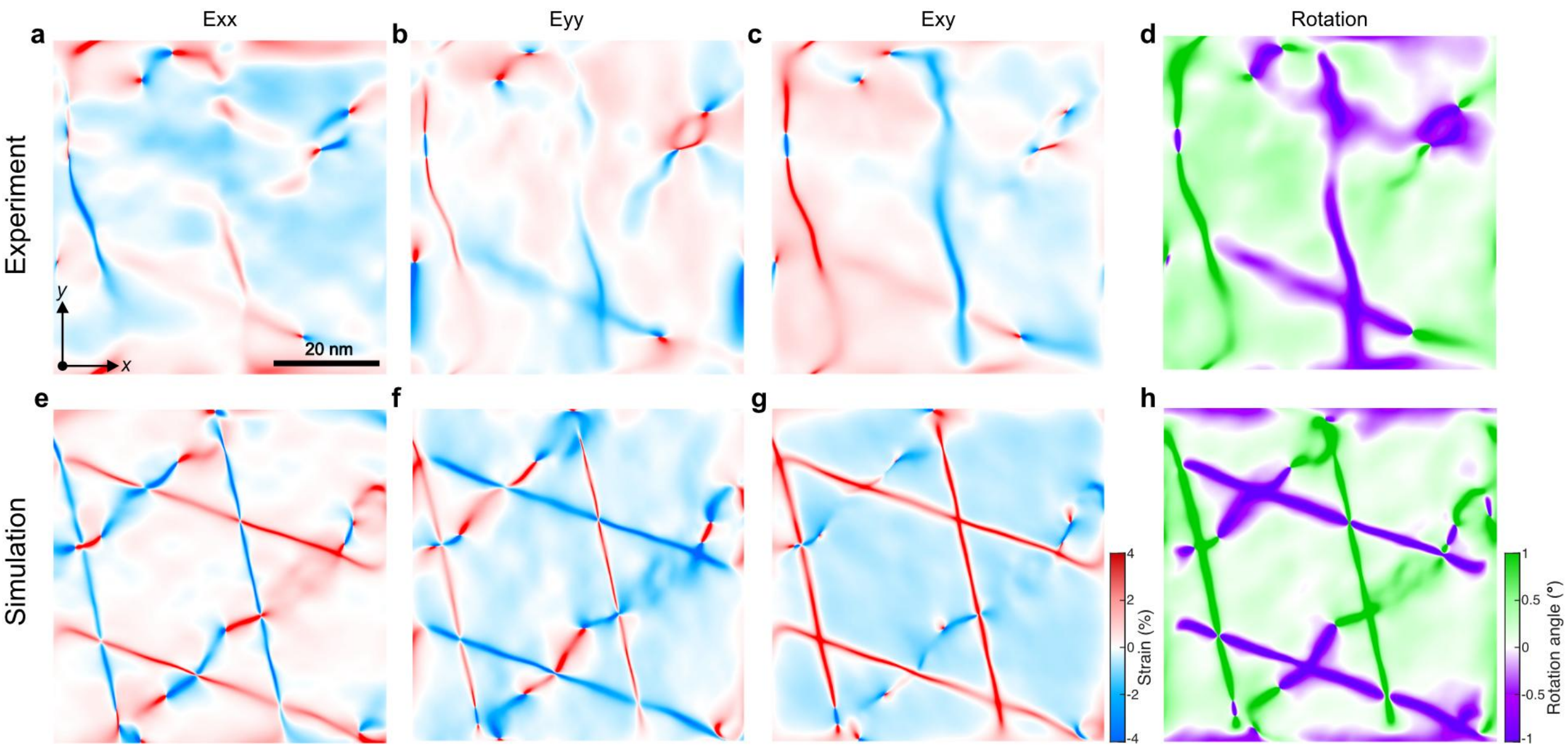

**Figure 4** Moiré distortion and rotation fields of Mt calculated by GPA of the focal series reconstruction image in Fig. 3 and simulated image. The distortion maps show three strain components: Exx (**a, e**), Eyy (**b, f**), and Exy (**c, g**). The rotation angle map (**d, h**) displays the in-plane twist angles. The simulated distortion maps and rotation angles were calculated from a 2D potential image of an ideal Mt bilayer atomic structure using GPA.

## Discussion

The combined experimental and simulation results establish a coherent physical picture for rotational disorder in strongly interacting layered materials. In Mt, small-angle misorientation does not arise from random stacking but from a competition between interlayer registry and electrostatic frustration that hydration strongly modulates. Because the energetic cost of twist is comparable to thermal and hydration energies, rotation becomes an accessible degree of freedom coupled to moiré formation, strain localization, and defect accommodation. Under nonequilibrium dehydration, twist configurations favored at higher hydration are kinetically inherited rather than fully relaxed, which naturally explains the dominance of small but finite angles.

The preference for finite, small-angle registries over perfect alignment at low hydration originates in the random isomorphic substitution of $Mg^{2+}$ for $Al^{3+}$ in the octahedral sheet. Because these substitutions are uncorrelated between adjacent layers, no single twist angle can optimize the registry of all charged sites, producing charge frustration analogous to that in spin glasses. Placing the same substitutions on an ordered sublattice removes the small-angle minima and yields a smooth, single-basin landscape (Fig. S3), confirming that charge disorder is the necessary ingredient. In this sense, glass denotes a metastable, kinetically trapped configuration drawn from many nearly degenerate twist states, lacking long-range rotational order yet retaining locally preferred registries frozen in during dehydration.

This electrostatically driven plasticity contrasts with weakly interacting van der Waals bilayers, where twist is largely geometric, and the landscape is flat. The focal series reconstruction and GPA maps reveal the real-space signature of this frozen configuration: long-wavelength moiré stripes in the shear and rotation fields, and registry-dependent coherence that varies across the bilayer. We show that Mt is fully exfoliated into individual layers in solution (Fig. S8), confirming that the moiré glass state forms upon drying.

The moiré glass framework may apply broadly to strongly interacting layered materials, including layered double hydroxides, charged oxide lamellae, graphene oxide, and other electrostatically active lamellar solids. More generally, rotational disorder in such systems is better viewed not as a stacking defect but as an emergent, history-dependent state governed by rugged energy landscapes and kinetic trapping, with consequences for mechanical response, ion transport, and chemical reactivity. Our control simulations addressed the 1W state. How charge order tunes the landscape across hydration states, and whether twist can be cycled by repeated (de)hydration, are natural next steps toward deterministic control of moiré structure in charged layered solids.

# Methods

### *Sample preparation*

Wyoming montmorillonite (SWy-2; Fig. S2, Table S1), obtained from the Source Clays Repository of The Clay Minerals Society, was used throughout this study. SWy-2 is a 2:1 dioctahedral smectite in which a central octahedral sheet is covalently linked on both sides to tetrahedral sheets through shared apical oxygen atoms, giving the idealized half-cell composition $M_x(Al,Mg,Fe)_2(Si,Al)_4O_{10}(OH)_2$, where M denotes exchangeable interlayer cations[31,42]. Isomorphic substitution of $Mg^{2+}$ (or $Fe^{2+/3+}$) for $Al^{3+}$ in the octahedral sheet and of $Al^{3+}$ for $Si^{4+}$ in the tetrahedral sheet generates a net negative layer charge that is compensated by these interlayer cations. An aqueous diluent phase (herein called clay saturated aqueous phase, CSAP), saturated with respect to SWy-2, was prepared by dialyzing SWy-2 (10 mg) against MilliQ water (1 L, resistivity > 18.2 MΩ·cm) for 1 week and filtering through a 0.02 µm filter (Whatman). Aqueous NaCl solutions were prepared from reagent grade salts in CSAP so that they remained equilibrated with respect to the clay, avoiding artifacts from clay dissolution.

### *Clay Treatment*

To obtain homoionic Mt, SWy-2 (2-10 g) was dispersed in NaCl (1 M, 50 mL made from a clay-saturated aqueous phase, CSAP) in a polypropylene bottle (50 or 1000 mL) and mixed on a rotating mixer (5 rpm) for seven days. Sedimentation of coarse particles was achieved by centrifugation (1000 rcf for 10 minutes). Fine particles were separated from the coarse sediment and transferred to a cellulose dialysis membrane (Spectra/Por 3, 3.5 kDa molecular weight cutoff). Excess salt in the supernatant was removed by dialysis against CSAP, which was replaced every day for seven days. The resulting clay gel was dried in a convection oven (60 °C for 14 hours, then 110 °C for 6 hours). Suspensions were prepared by redispersing homoionic powder in CSAP to a final concentration of 20 mg/mL and sonicating in an ultrasonic bath at 45°C for 24 hours.

### *High-resolution transmission electron microscopy*

Samples for electron microscopy were prepared by resuspending 5 mg dried SWy-2 powder in 1 mL CSAP by sonication for 1 hour in a 1.5 mL high-density polyethylene centrifuge tube. This solution was diluted by adding 100 µL to 900 µL of CSAP in a separate centrifuge tube, producing a 0.5 mg/mL solution. A second dilution step was repeated by adding 100 µL of the resulting solution to 900 µL of CSAP, producing a 0.05 mg/mL. A 200 mesh copper TEM grid with lacy carbon (Ted Pella) was plasma cleaned in Ar-21%$O_2$ for 10 seconds, dipped in CSAP, and 2 µL of the 0.05 mg/mL solution was deposited on the TEM grid. The grid was dried overnight at 125 °C before being put into the microscope.

HRTEM was acquired on the TEAM I microscope at the National Center for Electron Microscopy at Lawrence Berkeley National Laboratory under aberration-corrected imaging conditions at an acceleration voltage of 300 kV. Treated clay samples were deposited into TEM grids with lacy carbon films, dried under vacuum for 24 hours, and imaged using low-dose conditions to avoid beam damage. Mt is known to be electron-beam sensitive, and when exfoliated, does not generate strong phase contrast. To preserve structural integrity while enabling focal-series reconstruction, a total electron dose of approximately 100 $e$ $Å^{-2}$ was used under aberration-corrected imaging conditions. Images were recorded with a pixel size of 0.2 Å. For phase retrieval, a focal series consisting of 21 images was acquired at different defocus values. The initial defocus range spanned approximately from −33 to +47 nm with intervals of about 4 nm. These defocus values were subsequently refined during the focal-series reconstruction procedure through numerical optimization.

***Layer counting and misorientation analysis.***
The number of layers per particle was determined by directly counting individual layers in edge view HRTEM images (Fig. S1). Interlayer misorientation angles were obtained from fast Fourier transforms (FFTs) of plan view lattice images: for each bilayer region, the misorientation was measured as the angular separation between equivalent first order Bragg reflections arising from the two overlapping layers, and the measured values were binned to construct the distribution in Fig. 1e. In total, N = 52 particles were analyzed for the layer number distribution (Fig. 1d) and 233 interlayer interfaces for the misorientation distribution (Fig. 1e).

***MD simulations for misorientation-angle distribution of Mt bilayer***
Two parallel clay particles were generated using the experimentally determined crystal structure of Cs-montmorillonite[43] with equal edge lengths with the particle sizes of 26 nm. To generate permanently charged particles, we randomly substituted 15% of octahedral $Al^{3+}$ with $Mg^{2+}$ ions, yielding a surface charge density of −0.96 C $m^{-2}$, consistent with experimental observations for this high degree of isomorphic substitution[43]. The resulting clay charge was compensated by $Na^{+}$ counterions placed in fully hydrated configurations above the substitution sites on each clay particle. The particles were solvated at varying crystalline hydration states corresponding to one, two, and three water layers in the interlayer (1W–3W), with a 1 nm buffer of bulk solution surrounding the clay particles in all directions. Depending on the hydration state, the total system size ranged from approximately 355,000–425,000 atoms. We employed the OPC3 rigid three-point water model, which provides an improved description of polarizability, dielectric properties, ion hydration energies, and water-oxygen-to-ion distances relative to conventional three-site models[44,45]. Interatomic interactions involving clay atoms were described using ClayFF[46], while water-water, ion-ion, and ion-water interactions in solution were treated with the OPC3-optimized ion parameters from ref.[47].

To model the free energy of particle twisting, we employed constrained molecular dynamics within an umbrella sampling framework. The simulations were performed at discrete twist angles from 0° to 15° for each hydration state (1W, 2W, and 3W). For small twist angles (0–2°), a finer angular spacing of 0.5° was employed because the corresponding moiré periodicity becomes very large, requiring higher angular resolution to accurately capture variations in the free-energy landscape. The twist angle was defined as the relative rotation of the two parallel clay particles about the axis normal to their surfaces, with 0° representing perfect alignment. A harmonic biasing potential was applied along this rotational collective variable, with a force constant of 20 kcal $mol^{-1}deg^{-2}$ during an initial 1 ns equilibration phase, subsequently relaxed to 10 kcal $mol^{-1}deg^{-2}$ for the following 5 ns of production sampling. The stronger initial restraint ensured that the system was guided toward the target twist angle before sampling commenced under a softer constraint. The free energy profile (potential of mean force) was obtained by monitoring fluctuations in the restoring forces along the trajectory and integrating the constraint forces over the sampled twist angles, following a procedure analogous to our previous potential of mean force calculations[48,49].

The simulation pipeline consisted of initial structural relaxation using conjugate gradient energy minimization, followed by thermal equilibration in the NVT ensemble (2 ns, 298 K) and subsequent cell optimization in the NPT ensemble (5 ns, 298 K, 1 atm). The NPT stage employed the Berendsen barostat[50] with a pressure relaxation time of 0.1 ps solely for equilibration purposes; because this barostat does not rigorously sample the isobaric-isothermal ensemble, it was not used during the production phase. All free-energy calculations were carried out in the NVT ensemble. Temperature was controlled throughout using a Langevin thermostat[51] with a collision frequency of $\gamma = 1\ ps^{-1}$. Long-range electrostatic interactions were evaluated using the smooth particle-mesh Ewald method. All simulations were performed using the pmemd simulation engine from the AMBER molecular dynamics package, optimized for GPU architectures[52].

***Ordered-substitution control simulations.***

To isolate the role of charge disorder, we performed an additional set of free-energy calculations in which the 15% octahedral $Mg^{2+}$ for $Al^{3+}$ substitutions were arranged on a periodic (ordered) sublattice rather than placed randomly. The $Mg^{2+}$ substitutions were placed on a hexagonally ordered sublattice consistent with the intrinsic symmetry of the octahedral sheet, yielding a periodic substitution pattern with uniform site separation. Free energies were computed for the 1W state over 0–15° using the same biasing force constants and sampling times described above.

***Focal series reconstruction***

We performed Focal series reconstruction from the HRTEM focal series data using PhaseT3M[53], to obtain the projected potential of the misoriented Mt bilayer. The HRTEM focal series consists of 21 images acquired at different defocus values, with an initial pixel size of 0.2 Å. Reconstruction was carried out using 1000 iterations with binned images with an effective pixel size of 0.4 Å (binning 2). Because the sample thickness was less than 1 nm, a single-slice focal series reconstruction approach was applied.

***Generation of the Mt twisted bilayer and 2D projected potential***

We constructed an ideal bilayer Mt atomic model based on a single-layer structure in Fig. 1a-b. To reproduce the experimental observations, the top layer was rotated by ~0.8° to match the measured interlayer misorientation (Fig. 3), and an additional out-of-plane tilt about the [130] axis was introduced when generating Bragg-filtered images. A tilt angle of 7° provided the best agreement with the experimentally observed Bragg-filtered contrast (Fig. 3 and S5). From the structure, a 3D volume of atomic potential was calculated and then projected into a 2D potential to simulate the focal series reconstruction result. The projected potential was sampled with a pixel size of 0.4 Å. The atomic potential was derived by Fourier transforming the electron scattering factors[54]. The generation was performed using abTEM[55], employing 8 frozen phonon configurations to account for thermal vibrations.

***Boltzmann-weighted misorientation-angle distribution calculations***

Misorientation-angle-dependent Gibbs free energies were obtained from MD simulations for bilayer Mt systems 1W, 2W, and 3W intercalated water layers. To estimate the relative population of misorientation angles, we constructed a Boltzmann-weighted angular distribution based on the calculated Gibbs free energy. For each hydration state $i$, the relative probability was calculated as

$$P_i(\theta) \propto exp\left[-\frac{F_i(\theta)A}{k_B T}\right],$$

where $F$ is free-energy per unit area, $k_B$ is the Boltzmann constant, $A$ is an effective domain area, and $T$ is the simulation temperature (298 K). The total misorientation-angle distribution was obtained by combining contributions from the 1W, 2W, and 3W cases. For each candidate value of the effective domain area $A$, the relative contributions of the different hydration states were determined by minimizing the Earth Mover's Distance (EMD) between the simulated and experimental distributions. This procedure yields a single EMD value for each $A$. The optimal value of $A$ was then selected as the one that minimizes the EMD across all tested values. The resulting distribution was normalized such that the total probability sums to unity. For this optimized value of $A = 3.7\ nm^2$, the corresponding contributions of the 1W, 2W, and 3W cases are 0%, 48%, and 52%, respectively. Note that the optimization of the effective domain area $A$ and the corresponding contributions of the hydration states was performed using only the experimental misorientation angles at integer values (0°, 1°, 2°, 3°, …). Relative probabilities at integer misorientation angles (0°, 1°, 2°, 3°, …) were determined using the experimental histogram for the EMD-based optimization. Relative probabilities at selected small angles (0.5° and 1.5°) were calculated directly from MD simulations, whereas probabilities at other intermediate angles (e.g., 2.5°, 3.5°, …), shown as open circles (Fig. 2e), were obtained by interpolation of the MD-derived free-energy landscape and were used only to visualize the continuous distribution.

***Geometric phase analysis of moiré distortion fields***

Geometric phase analysis (GPA) was applied to focal-series-reconstructed phase images and to simulated projected potential images to quantify moiré-induced distortion and in-plane rotation fields. Because the Mt bilayer contains two layers that are rotationally misaligned by less than 1°, GPA was performed using effective reference peaks derived from the overlapping first-order Bragg reflections. Specifically, six first-order Bragg reflections corresponding to the <110>, <020>, and <1–10> lattice directions were used.

For each reflection, the local maximum of the overlapping peak was selected as the reference for phase-gradient calculation. This local maximum corresponds effectively to the average position of the two rotated Bragg peaks arising from the individual layers, thereby defining an effective reciprocal-space reference associated with the moiré superlattice rather than with a single atomic lattice. Phase gradients were calculated relative to this effective reference, yielding long-wavelength distortion and rotation fields that reflect moiré-induced modulation rather than atomic-scale elastic strain within an individual layer.

Reciprocal-space masks were chosen to encompass the overlapping Bragg peaks while excluding higher-order reflections and diffuse background. The masks were constructed using a generalized Lorentzian weighting function,

$$M(q) = \left(1 + \left(\frac{(q-q_0)^2}{R^2}\right)^2\right)^{-1},$$

where $q_0$ is the peak position. An aperture radius of $R$ = 0.0075 Å$^{-1}$ and a shape parameter of $p$ = 2 were used, providing a smooth weighting that balances spatial resolution and strain precision. The same GPA procedure, including peak selection, mask size, and phase-gradient calculation, was applied consistently to both experimental and simulated images to enable direct comparison. Distortion components (Exx, Eyy, Exy) and rotation angles were extracted following standard GPA formalism[41].

***Geometric phase analysis of layered-resolved strain fields using separated high-order reflections***

To obtain layer-resolved strain and rotation fields, GPA was additionally performed using well-separated high-order Bragg reflections. Unlike the moiré-based GPA described above, which employs an effective reference derived from overlapping first-order reflections, this analysis takes advantage of the fact that high-order diffraction peaks in the <330> family are sufficiently separated in reciprocal space at small interlayer rotation angles, allowing reflections from individual layers to be independently selected.

For this analysis, symmetry-equivalent <330> Bragg reflections associated with Layer 1 and Layer 2 in Fig. 3 were separately masked and used as reference peaks. Phase gradients were calculated relative to each layer-specific reference, yielding layer-resolved strain and in-plane rotation maps that reflect lattice-scale deformation within each individual layer rather than moiré-averaged modulation. Because the <330> reflections are well separated, this approach avoids averaging over contributions from the two layers and enables direct comparison of deformation fields between layers.

The same GPA workflow used for the moiré-based analysis was applied, including the use of the same generalized Lorentzian masks ($R$ = 0.0075 Å$^{-1}$, $p$ = 2). The masks were designed to isolate individual

Bragg peaks while excluding contributions from neighboring reflections. Strain components (Exx, Eyy, Exy) and rotation angles were extracted for each layer to facilitate comparison with the moiré-scale distortion fields.

***Static and dynamic light scattering.***

The Na-Mt suspension used for imaging was characterized on a goniometer-based LS Spectrometer (LS Instruments, Fribourg, Switzerland) equipped with a 638-nm fibre-coupled laser, with the sample held at 25 °C in an index-matched toluene bath. Static and dynamic data were collected over scattering angles of 10–150°, with the scattering vector defined as $q = (4\pi n/\lambda_0)\cdot\sin(\theta/2)$. In static light scattering, the high-q intensity followed a power law $I(q) \propto q^{-p}$ with $p = 1.98 \pm 0.02$, the Porod signature of thin two-dimensional sheets and thus of fully exfoliated layers. In dynamic light scattering, the relaxation rate from cumulant analysis scaled linearly with $q^2$, confirming purely translational diffusion and giving a diffusion coefficient $D = 2.21 \times 10^{-12}$ $m^2$ $s^{-1}$.

# Supplementary Information for "Hydration-controlled twist forms a moiré glass in charge-frustrated layered silicates"

Juhyeok Lee[1,2], Piotr Zarzycki[3], Colin Ophus[4], Jim Ciston[2], Benjamin Gilbert[3,5], Jillian F. Banfield[3,5], Mary C. Scott[2,6], Michael L. Whittaker[1,3*]

1. Materials Sciences Division, Lawrence Berkeley National Laboratory, Berkeley, CA 94720
2. National Center for Electron Microscopy, Molecular Foundry, Lawrence Berkeley National Laboratory, Berkeley, CA 94720
3. Energy Geosciences Division, Lawrence Berkeley National Laboratory, Berkeley, CA 94720
4. Department of Materials Science and Engineering, Stanford University, Stanford, CA 94305
5. Department of Earth and Planetary Science, University of California, Berkeley, Berkeley, CA 94720
6. Department of Materials Science and Engineering, University of California, Berkeley, Berkeley, CA 94720

*Corresponding author. email: mwhittaker@lbl.gov

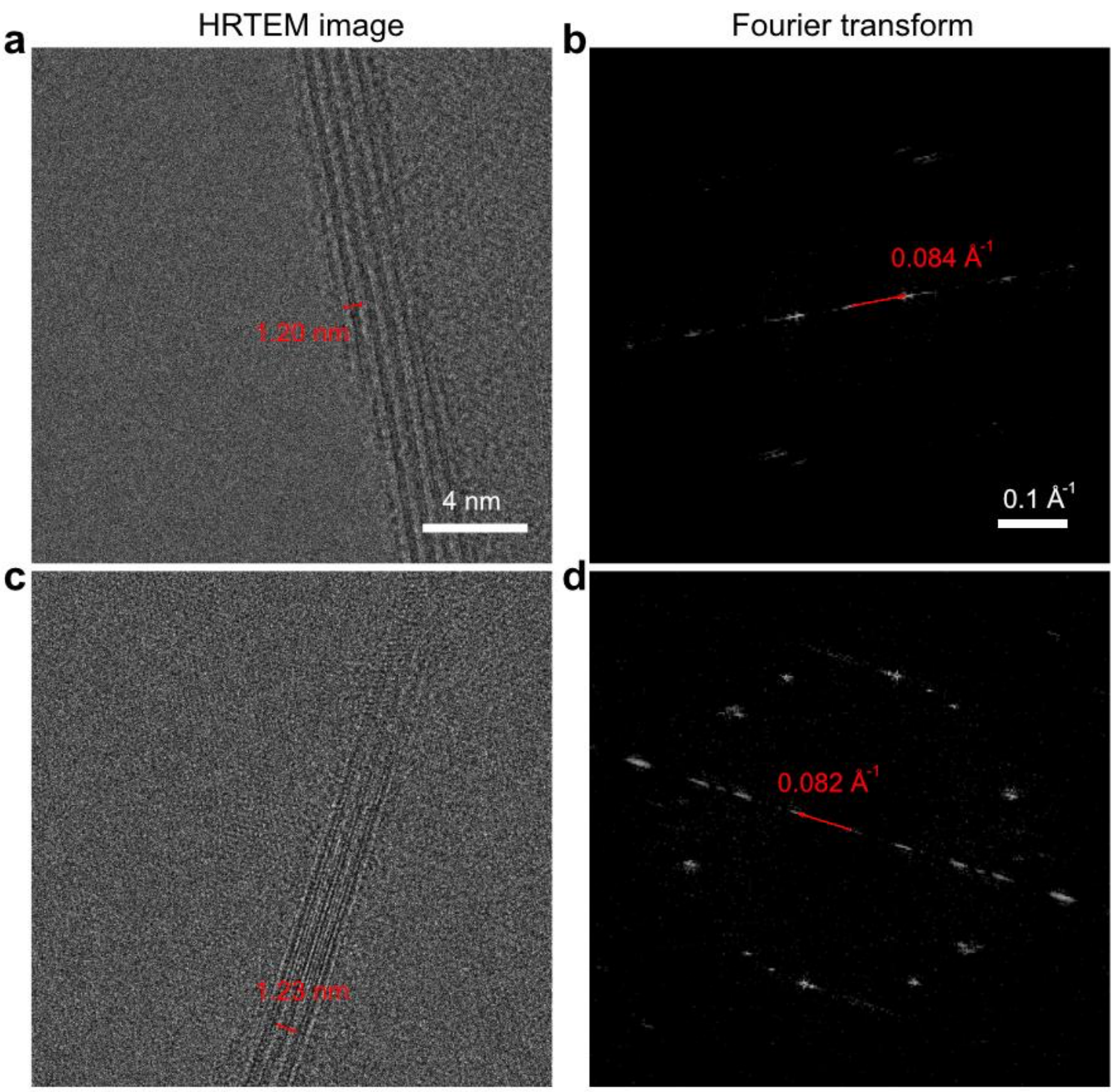

**Figure S1.** Edge-view HRTEM evidence for dehydration to the 1W hydration state during TEM measurement. (**a,c**) Representative edge-view HRTEM images of Mt bilayers acquired under the same imaging conditions. The measured interlayer spacings of ~1.20 nm (a) and ~1.23 nm (c) are consistent with the 1W hydration state. (**b,d**) Corresponding Fourier transforms of the images in (a) and (c), respectively, showing characteristic diffraction streaks associated with layered ordering. The indicated reciprocal-space spacings of ~0.084 Å$^{-1}$ (b) and ~0.082 Å$^{-1}$ (d) are consistent with the real-space interlayer periodicities measured in the HRTEM images.

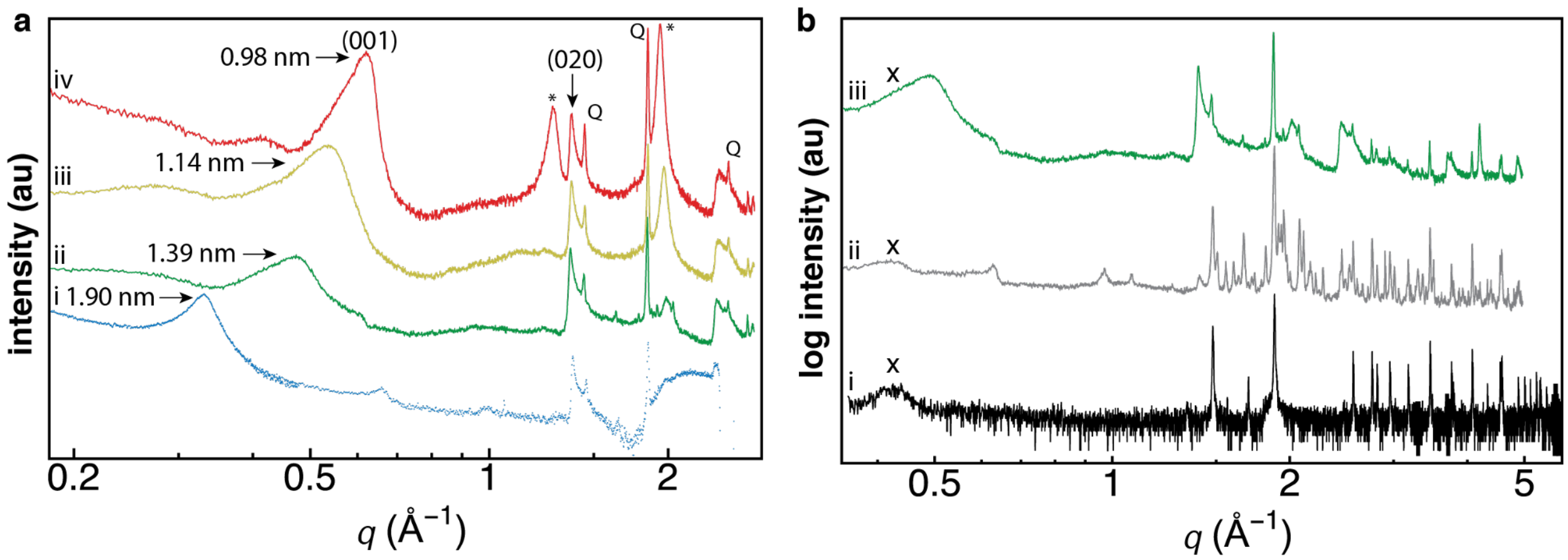

**Figure S2.** Powder XRD of montmorillonite. (**a**) WAXS of montmorillonite in 1 M NaCl (i) compared with pXRD of natural montmorillonite (ii), homoionized montmorillonite (iii), and size-fractionated montmorillonite (iv). Quartz peaks are labeled with (q), and (*) indicates harmonic peaks of the type (00*l*) where *l* = 4 or 6. (**b**) Quartz reference pXRD (i, from rruff.info) compared to the hard pellet separated during centrifugation treatment (ii) and natural SWy-2 (iii).

**Supplementary Table 1: peak indices, positions, and spacings from HRPXRD**

| peak index (*kh*0) | position ($g$, nm$^{-1}$) | spacing ($d$, Å) |
|---|---|---|
| 020 | 1.37 | 4.60 |
| 110 | 1.40 | 4.45 |
| 130 | 2.39 | 2.63 |
| 200 | 2.45 | 2.56 |
| 040 | 2.73 | 2.30 |
| 220 | 2.80 | 2.24 |
| 150 | 3.63 | 1.73 |
| 240 | 3.67 | 1.71 |
| 310 | 3.73 | 1.68 |
| 060 | 4.10 | 1.53 |
| 330 | 4.20 | 1.50 |
| 360 | 4.77 | 1.32 |
| 170 | 4.94 | 1.27 |
| 350 | 5.01 | 1.25 |
| 420 | 5.08 | 1.24 |
| 260 | 5.10 | 1.22 |

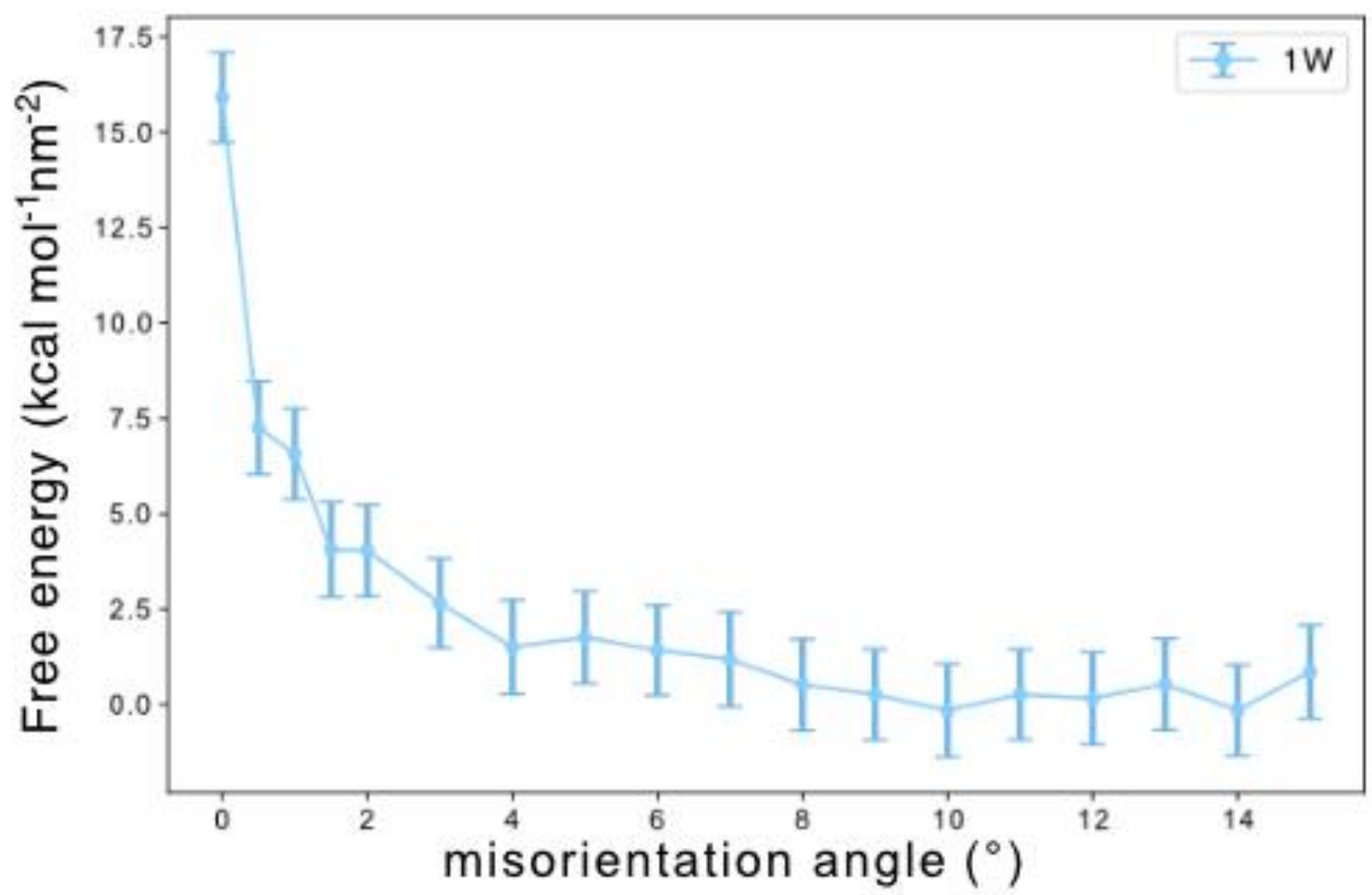

**Figure S3. Misorientation-angle-dependent free energy of an Mt bilayer with ordered $Mg^{2+}$ substitution (1W state).** Relative Gibbs free energy per unit area versus interlayer misorientation angle for a bilayer in which the octahedral $Mg^{2+}$ for $Al^{3+}$ substitutions are placed on a regular (ordered) sublattice rather than randomly, computed with the same umbrella-sampling protocol as in Fig. 2 (identical charge density, particle size, water model, and counterion content). Free energies are shown relative to the minimum value; error bars represent root-mean-square fluctuations from MD sampling. In contrast to the disordered case (Fig. 2d), the ordered arrangement produces a smooth, single-basin landscape that decreases nearly monotonically from a maximum at perfect alignment (0°) to a broad shallow minimum at large angle, without the small-angle local minima (~1–2°) that dominate the disordered landscape. The absence of these metastable wells indicates that the rugged free-energy landscape underlying the moiré glass originates from charge disorder rather than from interlayer twist itself.

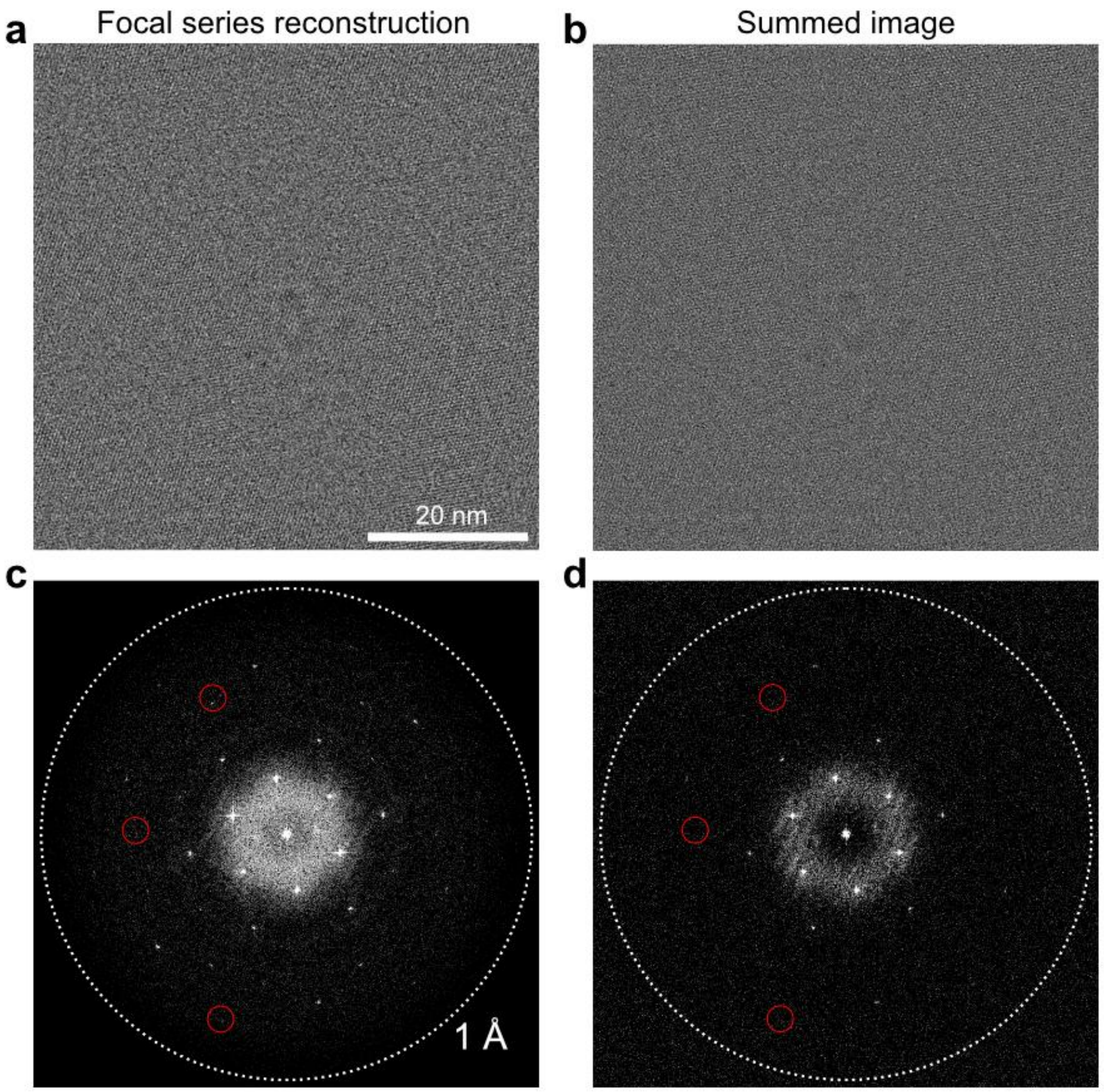

**Figure S4.** Comparison between focal series reconstruction (**a,c**) and summed HRTEM image (**b,d**) of Mt bilayer. The focal series reconstruction provides enhanced resolution, resolving moiré domains and local lattice alignments that are less apparent in the summed image.

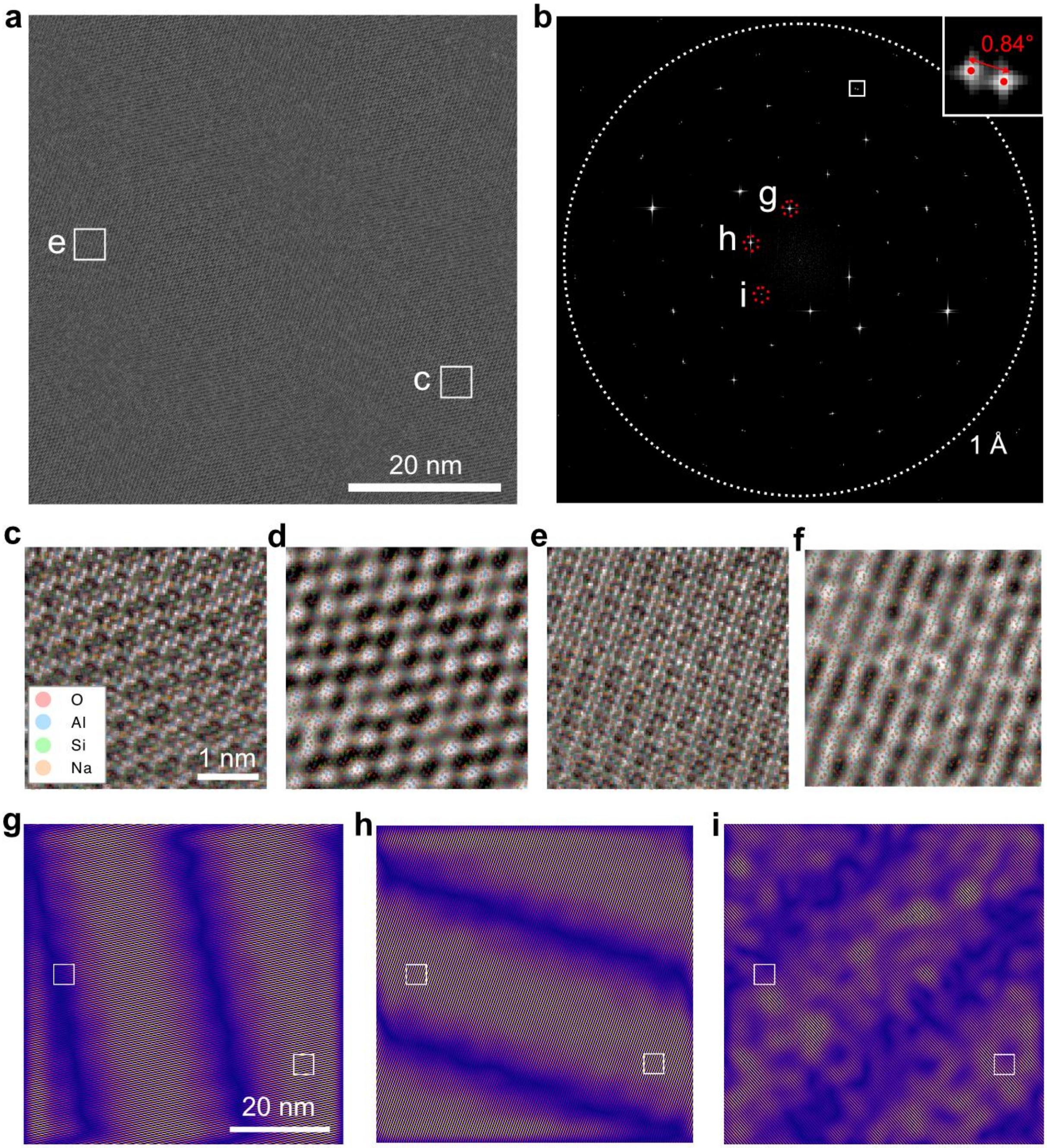

**Figure S5.** Simulated 2D projected image from ideal Mt bilayer Moiré with 0.84° misorientation with a 7° global tilt about the [130] tilt axis. (**a**) Overview image. (**b**) Fourier transform of (**a**), revealing (260) peaks split by 0.84°. (**c-f**) Enlarged regions from (**a**), showing (**c**) alignment along all three principal directions, (**e**) alignment along (020) only. A low-pass filter with a cutoff of 0.3 $Å^{-1}$ is applied to the enlarged images (**c,e**), yielding the filtered images (**d,f**), respectively. (**g-i**) Bragg filtered images from (110) (**g**), (020) (**h**), and (1-10) (**i**) overlaid on top of (**a**). In contrast to the no-tilt case (Fig. S6), the inclusion of global tilt yields improved agreement with the experimental images, reproducing the anisotropic distortion in the (1–10) direction as well as the hexagonal cluster-like features and stripe patterns observed in Fig. 3b,d,i.

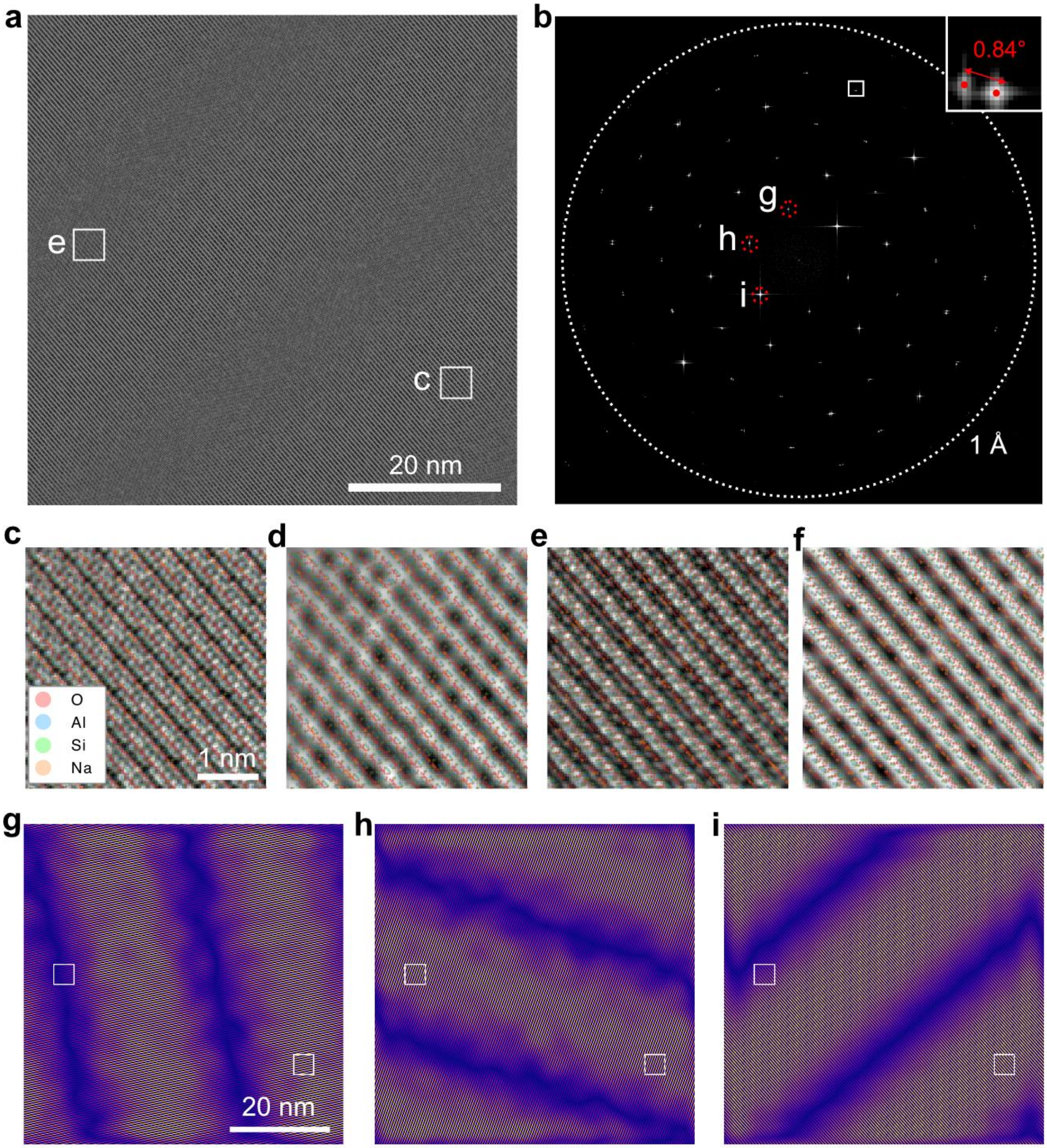

**Figure S6.** Simulated 2D projected image from ideal Mt bilayer Moiré with 0.84° misorientation in the absence of global tilt. (**a**) Overview image. (**b**) Fourier transform of (**a**), revealing (260) peaks split by 0.84°. (**c-f**) Enlarged regions from (**a**), showing (**c**) alignment along all three principal directions, (**e**) alignment along (020) only. A low-pass filter with a cutoff of 0.3 $Å^{-1}$ is applied to the enlarged images (**c,e**), yielding the filtered images (**d,f**), respectively. (**g-i**) Bragg filtered images from (110) (**g**), (020) (**h**), and (1-10) (**i**) overlaid on top of (**a**). In the absence of global tilt, the simulated Bragg-filtered image in (**i**) does not reproduce the anisotropic distortion observed in Fig. 3i. Likewise, the filtered images (**d**,**f**) do not capture the hexagonal cluster-like features and stripe patterns observed in Fig. 3b,d, indicating that global tilt is required to reproduce these experimental features.

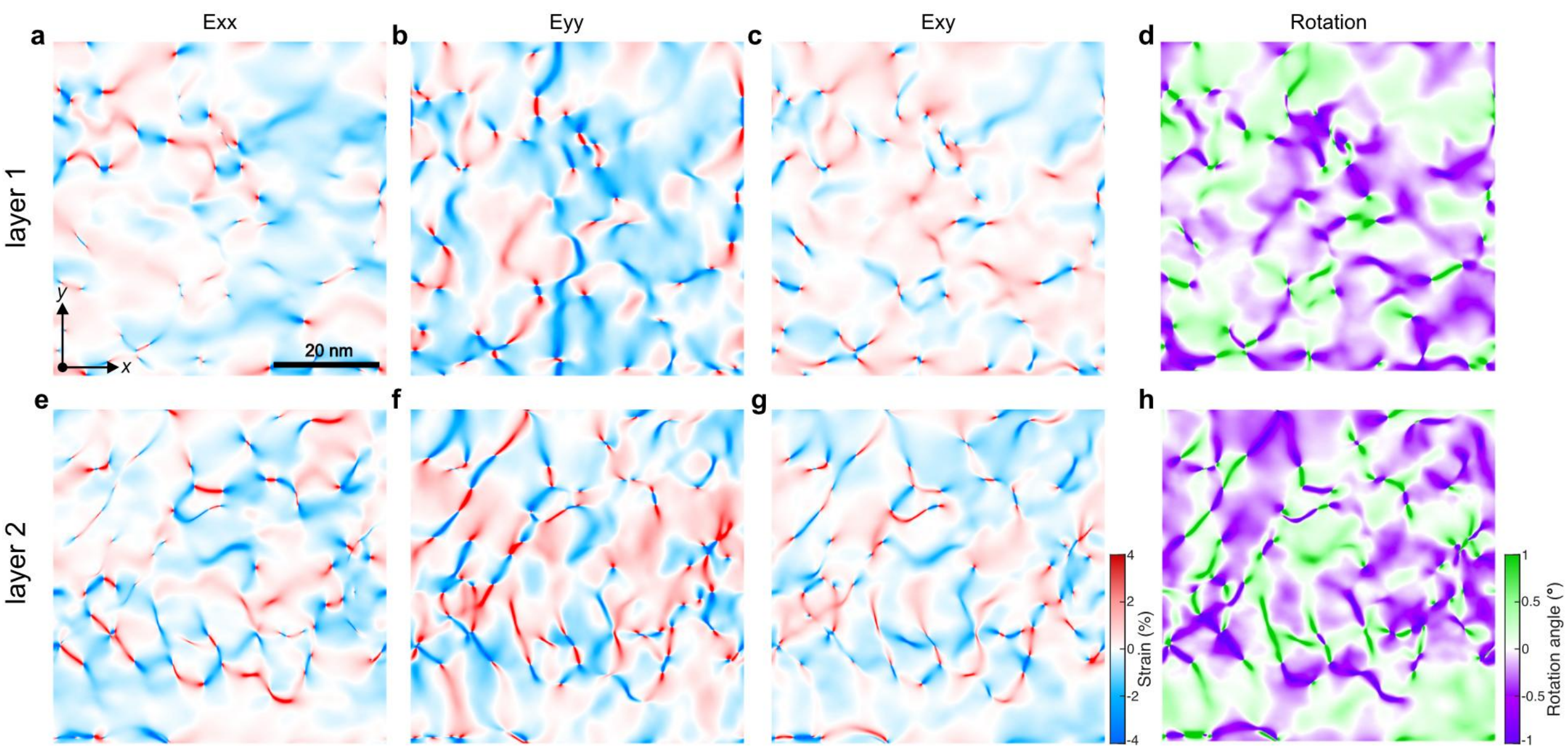

**Figure S7.** Layer-resolved strain and rotation maps of an Mt bilayer obtained from geometric phase analysis. Strain maps and in-plane rotation angles were calculated by GPA of the focal-series-reconstructed image in Fig. 3, using well-separated high-order <330> Bragg reflections corresponding to Layer 1 and Layer 2 of the Mt bilayer. The GPA maps show three components of the effective strain field for each layer: Exx (a, e), Eyy (b, f), and Exy (c, g). The rotation angle maps (d, h) display the in-plane twist angles of Layer 1 and Layer 2, respectively.

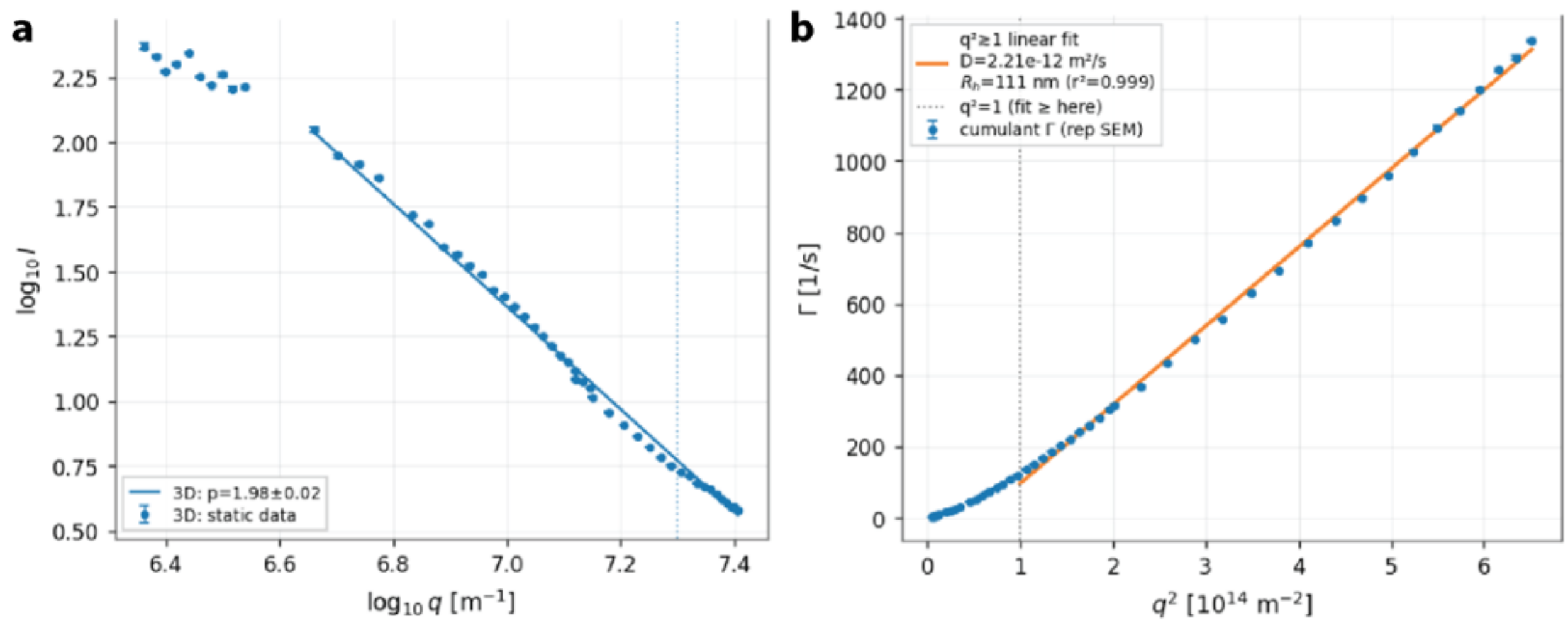

**Figure S8. Static and Dynamic light scattering of Mt suspension prior to deposition for imaging.** (a) Static light scattering exhibits primarily Porod scattering with a slope of $p$=1.98 ± 0.02, which is consistent with fully exfoliated Mt layers. Guinier scattering at smaller scattering vectors, $q$, indicates the presence of aggregated flocs of weakly-interacting layers with characteristic dimensions on the order of microns. (b) Angle-resolved dynamic light scattering indicates an exfoliated layer diffusivity of $D = 2.21 \times 10^{-12}$ m$^2$/s.